\documentclass[preprint]{revtex4-1} 
\usepackage{graphicx}
\usepackage{bm}
\usepackage{amsmath}
\usepackage{dcolumn}
\usepackage{hyperref}
\usepackage{natbib}
\begin{document}
\title{Two-dimensional coherent photocurrent excitation spectroscopy in a polymer solar cell}
\date{\today}
\author{Eleonora~Vella, Pascal~Gr\'egoire}
\affiliation{D\'epartement de physique \& Regroupement qu\'eb\'ecois sur les mat\'eriaux de pointe, Universit\'e de Montr\'eal, C.P.\ 6128, Succursale centre-ville, Montr\'eal (Qu\'ebec) H3C~3J7, Canada}  
\author{Hao~Li}
\affiliation{Departments of Chemistry and Physics, University of Houston, Houston, Texas 77204, USA}
\author{Sachetan~M.~Tuladhar, Michelle~Vezie, Sheridan~Few, Jenny~Nelson} 
\affiliation{Department of Physics, Imperial College London, South Kensington Campus, London SW7~2AZ, United Kingdom}
\author{Eric~R.~Bittner}
\affiliation{Departments of Chemistry and Physics, University of Houston, Houston, Texas 77204, USA}
\author{Carlos~Silva}
\affiliation{D\'epartement de physique \& Regroupement qu\'eb\'ecois sur les mat\'eriaux de pointe, Universit\'e de Montr\'eal, C.P.\ 6128, Succursale centre-ville, Montr\'eal (Qu\'ebec) H3C~3J7, Canada}  

\begin{abstract}
In high-performance solar cells based on polymeric semiconductors, the mechanism of photocarrier generation on $<100$-fs timescales is yet to be unravelled. In particular the dynamics of early-time electronic coupling between excitons on polymer chains and charge-transfer states need to be investigated in order to develop a detailed picture of ultrafast processes involved in photocurrent production. In this proceeding, we report preliminary measurements using a novel spectroscopy that can measure such correlations: two-dimensional coherent photocurrent excitation spectroscopy. This nonlinear technique measures off-diagonal spectral correlations in a two-dimensional photocurrent excitation spectrum. We interpret these spectroscopic measurements in light of recent theoretical predictions. 
\end{abstract}
\keywords{Organic solar cells; two-dimensional photocurrent excitation spectroscopy; photocarrier dynamics; polymeric semiconductors.}
\maketitle   
\section{Introduction}
An important challenge in the realization of efficient molecular-based solar cells lies in the understanding of efficient charge photogeneration mechanisms on ultrafast timescales. Organic photovoltaic devices (OPVs) based on the so-called bulk heterojunction consist of blends of electron donor (usually a $\pi$-conjugated polymer) and electron acceptor (typically a fullerene derivative or a second polymer) materials~\cite{Halls:1995uq}.
In such systems, the primary photoexcitations are tightly bound intrachain excitons with high binding energy ($\sim0.5$\,eV). Furthermore, the binding energies of the lowest-energy charge-separated states that are produced by ultrafast photoinduced charge transfer have comparable values~\cite{Gelinas:2011vn}. This implies that photocarrier generation is an energetically expensive process in these materials. 
Nonetheless, in the best OPV blends, device internal quantum efficiencies can approach unity~\cite{Park:2009ys}.
Within this framework, the ultrafast photocarrier generation dynamics in OPVs is a fundamentally unresolved issue~\cite{FalkeScience2014,VandewalNatMat2014,GelinasScience2014,ProvencherNatComms2014}.
It has been considered for some time that Coulomb-bound electron-hole pairs pinned to the interface are precursors to photocarriers~\cite{Clarke:2010fk}. Nevertheless, various reports indicate that unbound charges are produced on $<100$-fs timescales~\cite{GelinasScience2014,ProvencherNatComms2014}, bringing the role of tightly-bound charge-transfer species under question. In Ref.~\cite{GelinasScience2014}, by analysis of electroabsorption signatures in transient absorption spectra, the separation of the photogenerated electron-hole pairs was determined to be at least 4\,nm within the first 40$\,$fs after the excitation, implying a mean residual electron-hole Coulomb attraction near or below thermal energies. On the other hand, in Ref.~\cite{ProvencherNatComms2014} femtosecond stimulated Raman spectroscopy was used to probe the hole-polaron formation dynamics on the polymer. The data were interpreted to suggest that polarons, emerge directly on $<100$-fs timescales, and that these are free from the geminate electron on the fullerene, suggesting an electron-hole separation in the order of several nanometers over ultrafast timescales.
Recent work, based on both theoretical and experimental results~\cite{GelinasScience2014,BittnerNatComms2014,TamuraJACS2014} have invoked a coherent nature of delocalized charges and excitons in the charge separation process. In particular by means of a fully quantum mechanical/finite temperature model for the polymer heterojunction based on the configuration interaction theory, some of us reported on the role of the noisy environment in inducing transitions between excitons localized on a polymer chain and delocalized charge-separated states, allowing for the hole and the electron to be sufficiently far apart across the heterojunction so that charge carriers can be readily produced~\cite{BittnerNatComms2014}.  These reports, whether proposing the involvement of CT states~\cite{TamuraJACS2014} or of polarons~\cite{GelinasScience2014,BittnerNatComms2014} in the early stages of the charge transfer process, all support the notion of photocurrent generation via access to delocalized states. Addressing this scientific problem then requires the employment of an experimental probe of off-diagonal couplings between the exciton states localized on the polymer chain and delocalized $\pi$-electron states of the acceptor material. 

Here we introduce our implementation of a novel nonlinear photocurrent excitation spectroscopy, two-dimensional photocurrent excitation spectroscopy (2DPC), in which spectral correlations can be identified and time-resolved with femtosecond resolution in a two-dimensional spectrum, and we discuss its employment to unravel the charge generation process in OPVs. In general in a 2D coherent spectroscopic measurement, the non-linear response of a given system that has been excited by a sequence of phased ultrafast laser pulses is recorded in a multidimensional frequency or time space. By picking up the desired non-linear signal with the appropriate phase relationship one determines the superposition of states induced by the interaction with the exciting fields at separate instants of time. 2D coherent spectroscopy is particularly appealing because it exposes directly the existence of couplings between distinct excited states. 
In the context of photocarrier generation dynamics, this property is key for the investigation of energy and charge transfer phenomena that eventually lead to photocurrent. 

\section{Experimental}
\subsection{2D-PCS  apparatus}
\begin{figure}%
\includegraphics[scale=0.3]{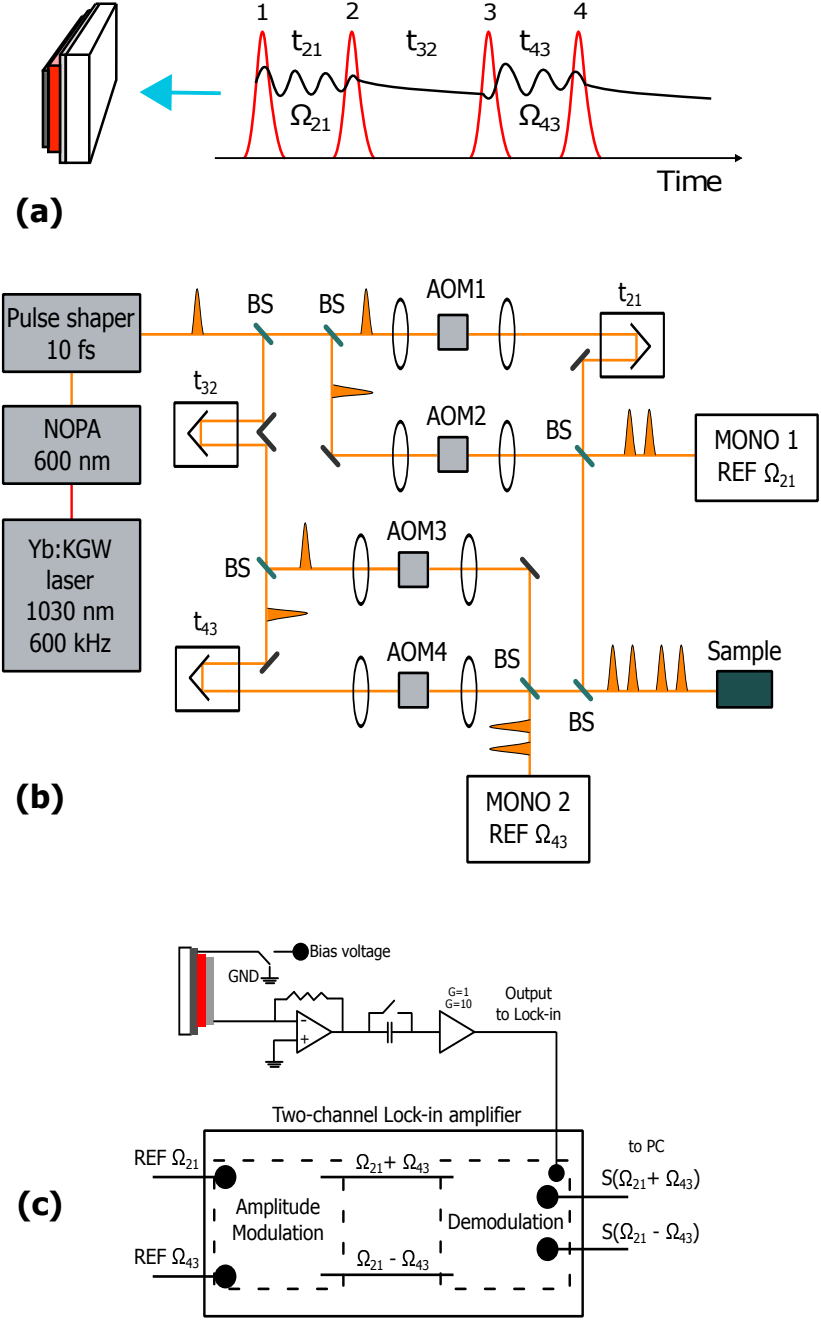}%
  \caption{ \label{fgr:setup}(a) Pulse sequence used in the experiment. (b) Schematic diagram of the set-up for the bi-dimensional spectroscopy experiment in collinear geometry. Abbreviations have the following meanings: NOPA, non-linear optical parametric amplifier; BS, beam splitter; AOM, acousto-optical Bragg cell; MONO, monochromator; REF, reference. (c) Schematics of the apparatus for the photocurrent detection.}
\end{figure}

Our 2D-PCS set-up is based on four collinear femtosecond pulses and acousto-optic phase modulation~\cite{TekavecJCP2007}.The four collinear pulses are generated using the apparatus schematically shown in Fig.~\ref{fgr:setup}(b). The output pulse train of a regenerative amplifier (light Conversion Pharos, $\lambda_0$=1030$\,$nm, pulse duration $\sim 200$\,fs) operated at a repetition rate of 600\,kHz, pumps a Non-collinear Optical Parametric Amplifier (NOPA) delivering large bandwidth pulses ($\sim 80$ -- 100\,nm) centred at around 600\,nm. The pulses are compressed by an adaptive $4f$ pulse shaper (BioPhotonics Solutions FemtoJock-P) in order to be $\sim 10$-fs long at the sample position. By mean of a 50\% beam-splitter (BS) the pulse train is then split to feed two twin Mach-Zehnder interferometers nested in a larger Mach-Zehnder interferometer. Three computer controlled delay stages allow independent control of the three interpulse delays $t_{21}$, $t_{43}$ and $t_{32}$ (see Fig.~\ref{fgr:setup}(a)).  Each of the twin Mach-Zehnder interferometers produces two of the phase-locked pulses. The phase-locking condition is achieved by placing in each arm of the two twin interferometers an acousto-optic Bragg cell which modulates each pulse at an unique frequency ($\Omega_i$, $i=1$, 2, 3 and 4) close to 200$\,$MHz. The exact $\Omega_i$ values are chosen so that the difference between the modulation frequencies of the two pulses generated by each of the twin Mach-Zehnder interferometers (indicated in Fig.~\ref{fgr:setup}(b) as $\Omega_{21}$ and $\Omega_{43}$) is of the order of $\sim 10$\,kHz. 
%%%%%%%%%%%%%%%%%%%%%%%%%%%%%%%%%%%%%%%%%%%%%
%%%%%%%%%%%%%%%%%%%%%%%%%%%%%%%%%%%%%%%%%%%%%%
For a detailed description of the phase modulation scheme adopted in the experiment, we refer the readers to Ref.~\cite{TekavecJCP2007} (see also Ref.~\cite{NardinOE2013} for a related approach), here we will briefly outline its basic principle. The time interval between consecutive laser pulses ($T_{\textrm{rep}}$) is set by the laser repetition rate, 600$\,$kHz. Due to the acousto-optic modulators, each of the pulses generated by the two twin Mach-Zehnder interferometers has a frequency shift imparted, equal to the respective frequency $\Omega_i$. Although these frequency changes are negligeable as compared to the optical frequency, 
they introduce a shift in the temporal phase of each pulse that oscillate at the corresponding frequency $\Omega_i$. 
At each laser shot, the experiment is repeated with a sequence of pulses phase-shifted with respect to the previous shot, thus creating two collinear trains of phase-modulated pulse pairs. Those interfering excitation pulses produce a population signal oscillating at $\Omega_{21}$ and $\Omega_{43}$ in the kHz range. It can be readily shown that~\cite{TekavecJCP2007}, within this phase cycling scheme, the non-linear signals of interest, which in analogy to four wave mixing experiments are referred to as \textit{rephasing} and \textit{non-rephasing} signals, oscillate at the frequencies $\Omega_{43}-\Omega_{21}$ and $\Omega_{43}+\Omega_{21}$ respectively. These frequency components are extracted simultaneously from the overall photocurrent signal by dual lock-in detection. The phase-sensitive detection scheme also offers a dinstinct advantage in the self-stabilization of the experiment. As it is possible to see in Fig.~\ref{fgr:setup}(b), optical replicas of pulses 1, 2 and 3, 4 are generated at the exit beam splitters of the two twin Mach-Zehnder interferometers and used to generate the reference signals for the dual lock-in demodulation. The two sets of pulse replicas are sent to two monochromators, which spectrally narrow them (thus temporally elongating them), and are then detected by two avalanche photodiodes. The temporal elongation of the pulses provided by the monochromators produces reference signals for time delays $t_{21}$ and $t_{43}$ up to $\sim10$\,ps. It is important to note that, when scanning $t_{21}$, the phase of the references built in this way does not evolve at an optical frequency, but at a reduced frequency given by the difference between the frequency of the signal and that set by the monochromators. This frequency downshift results in an improvement of the signal-to-noise ratio, which is inversely proportional to the frequency downshift itself, and which virtually removes the impact of the mechanical fluctuations occurring in the setup on the signals of interest~\cite{TekavecJCP2007}. In order to obtain the reference signals at the frequencies of the rephasing and non-rephasing signals, $\Omega_{43}-\Omega_{21}$ and $\Omega_{43}+\Omega_{21}$, one of the two photodiode outputs (typically the one at higher frequency) undergoes amplitude modulation (AM) by the output of the other photodiode. The AM signal so obtained carries the two sideband frequencies of interest ($\Omega_{43}-\Omega_{21}$ and $\Omega_{43}+\Omega_{21}$) and can then be used for the lock-in demodulation, as schematically shown in Fig.~\ref{fgr:setup}(c).

The 2D maps are built acquiring the demodulated signals at fixed $t_{32}$ and by scanning $t_{21}$ and $t_{43}$; $t_{21}$ is called the coherence time, $t_{43}$ the detection time and $t_{32}$ the population waiting time. Specifically, for a given detection time, data are sequentially recorded at different coherence times in the interval of interest, typically extending to a few hundred femtoseconds, the detection time is then stepped, and the coherence time scan repeated until the full 2D time response is recorded. Each of such scans simultaneously produces four maps: the in-phase and the in-quadrature ones for the rephasing and non-rephasing frequencies. The maps so obtained in the time domain are, finally, converted to the energy domain by Fourier-transforming the time variables $t_{21}$ and $t_{43}$ and recorded as a function of the population waiting time, $t_{32}$.

\subsection{Materials and sample preparation}
The devices examined in the present work are based on blends of poly(N-90-heptadecanyl-2,7-carbazole-alt-5,5-(40,70- di-2-thienyl-20 ,10 ,30 -benzothiadiazole)) as electron donor and of the fullerene derivative [6,6]-phenyl-C60 butyric acid methyl ester as electron acceptor (PCDTBT:PC60BM). 
The PV device structure is ITO/PEDOT:PSS/PCDTBT: PC60BM/Ca/Al. PEDOT:PSS was spun onto a cleaned ITO coated glass substrate to form a film of ~35$\,$nm thickness. The active layer of thickness ~80-90$\,$nm was then spin cast on top of PEDOT:PSS layer from a blend of PCDTBT:PC60BM (1:2) in chlorobenzene solvent with a concentration of 25mg/ml. The top electrode calcium (~20$\,$nm) and aluminium (~100$\,$nm) was then subsequently deposited by thermal evaporation. Current densityÐvoltage (JÐV) characteristics of the devices were measured using a Keithley 236 Source Measure Unit. Solar cell performance was measured by using a xenon lamp with AM1.5G filters and 100$\,$mW/cm$^{-2}$ illumination  solar simulator (Oriel Instruments) .
\subsection{Theoretical modeling}
Computer simulations were performed using parameters derived from the heterojunction lattice model described in Ref.~\cite{BittnerNatComms2014}. This two-band exciton/phonon model is parametrized to describe the $\pi$-electronic states of a polymeric semiconductor heterojunction and it includes both stacking and a band offset to define a heterojunction domain. The photoexcitations at the heterojunction are described by a four-level system involving the ground and the bound CT states, the polaron state being responsible for the photocurrent, as well as the excitonic state which corresponds to the only optically allowed excitation in the model. The energies of the different states are: -2.1\,eV for the exciton, -1.8\,eV  for the CT state and -2.2\,eV for the polaron state. State tunneling is allowed between the three excited states. The exciton-polaron correlation (0.1 eV) is set to be five times stronger than the exciton-CT one (0.02 eV), whereas the tunneling between CT and polaron states are prohibited. The dephasing effect is described by the Lindblad master equation and it relies on the trace-preserving and completely positive form of the reduced density matrix. The environmental noise is considered by the bath on the system resulting dephasing effect, which is attributed to state-energy oscillation and the transfer integral modulation. In the results shown here, the system decay is not taken into account. In other words, the system of interest is conserved in both energy and the number of quasiparticles.

\section{Results and discussion}
%%%%%%%%%%%%%%%%%%%%%%%%%%%%%%%%%%%%%%%%%%%%%
\begin{figure}%
\includegraphics[scale=0.5]{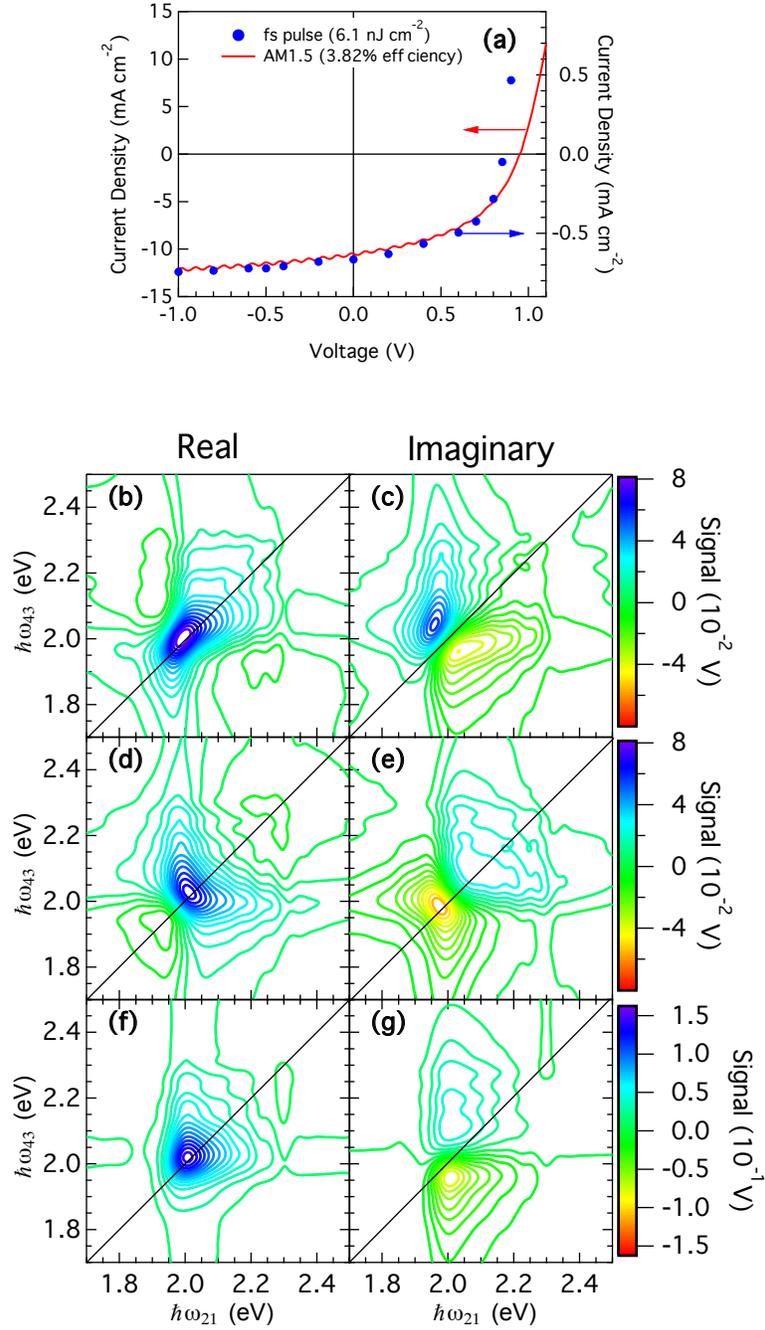}
\caption{\label{fgr:2DPC}(a) J-V characteristics of the PCDTBT:PC60BM photocell under solar simulator illumination (solid line) and under pulsed fs excitation. Photocurrent detected 2D spectra on the same device: (b)-(c) rephasing signal, (d)-(e) non-rephasing signal and (f)-(g) total correlation function. All spectra were recorded at a population waiting time of 50$\,$fs.}
\end{figure}
%%%%%%%%%%%%%%%%%%%%%%%%%%%%%%%%%%%%%%%%%%%%%
%%%%%%%%%%%%%%%%%%%%%%%%%%%%%%%%%%%%%%%%%%%%%
\begin{figure}%
\includegraphics[scale=0.5]{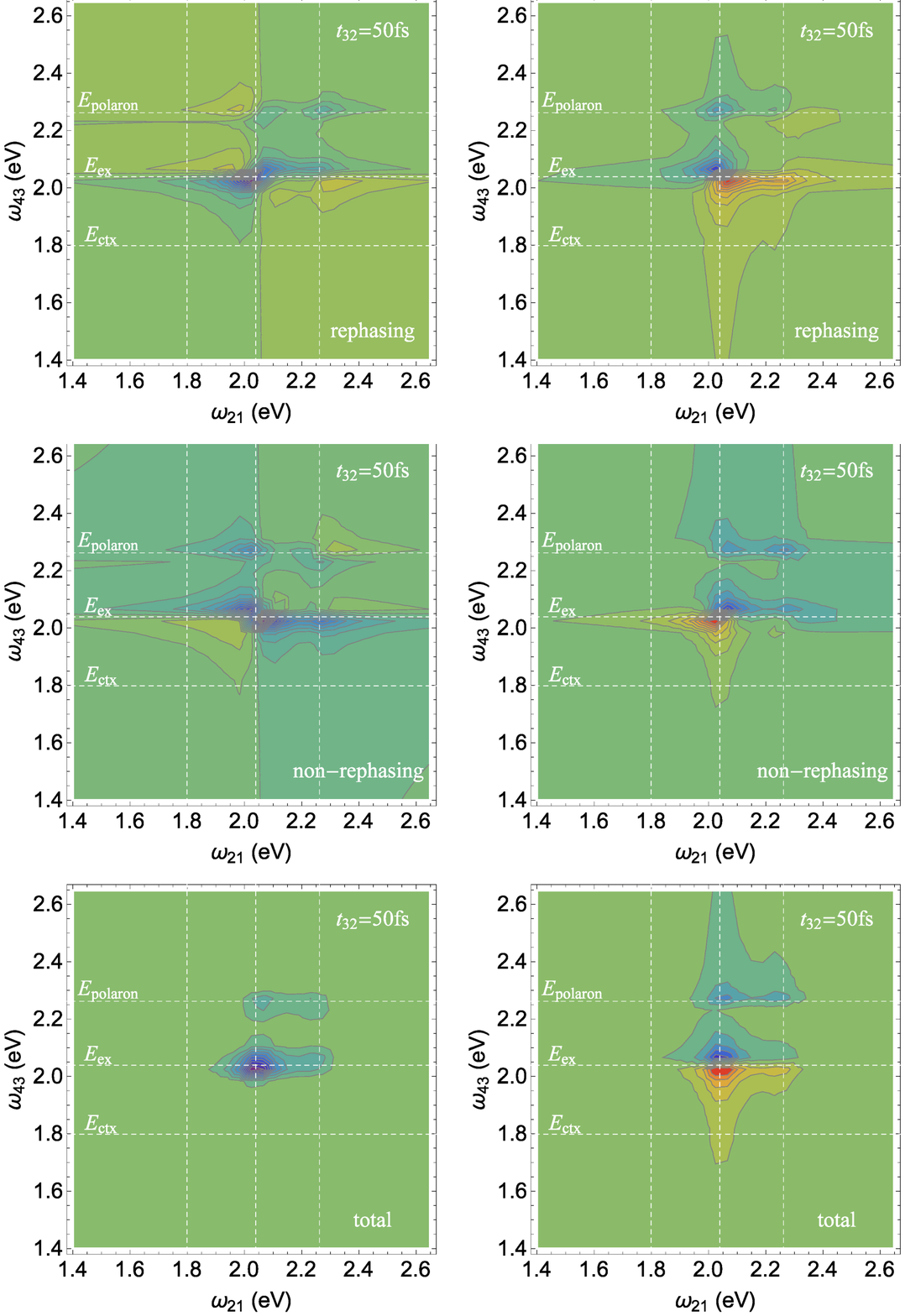}
\caption{\label{fgr:2DPCmodelling} Modeled rephasing, non-rephasing and total correlation function 2D spectra for a population waiting time of 50$\,$fs.}
\end{figure}
%%%%%%%%%%%%%%%%%%%%%%%%%%%%%%%%%%%%%%%%%%%%%
In fig.~\ref{fgr:2DPC}(a) the J-V characteristics on the PCDTBT:PC60BM blend solar cell are reported as measured under illumination with a solar simulator and under pulsed fs excitation. In the latter case the excitation conditions were the same as in the 2D spectra described in the following. The curves shown in fig~\ref{fgr:2DPC}(a) proves that the 2D spectroscopy experiment was carried out in an excitation regime comparable to the actual working conditions of the photocell.  Fig.~\ref{fgr:2DPC} shows the 2DPC real and imaginary spectra on the photocell. The rephasing (Fig.~\ref{fgr:2DPC}(b)-(c)) and non-rephasing (Fig.~\ref{fgr:2DPC}(d)-(e)) spectra were recorded simultaneously using a two-channel lock-in amplifier. The total correlation function maps (Fig.~\ref{fgr:2DPC}(f)-(g)) are also displayed. A total correlation spectrum is a common way to display 2D spectroscopic data as it results from the sum of the rephasing and non-rephasing maps. All spectra were acquired at a population waiting time $t_{32}=50$\,fs.

Here, we now focus on the total correlation function maps of Fig.~\ref{fgr:2DPC}(f)-(g). The real part shows a primarily absorptive line-shape and it presents an asymmetric profile with a high-energy tail. In particular an off-diagonal structure at $\sim100$\,meV above the diagonal signal is visible under the broad main spectral feature. In a previous publication based on a heterojunction lattice model, described in Ref.~\cite{BittnerNatComms2014}, Bittner and Silva proposed that due to the spectral density derived from fluctuations due to coupling of the system with the noisy bath, off-diagonal coupling between various asymptotic states is possible, resulting in resonant tunnelling between them. In Fig.~\ref{fgr:2DPCmodelling}, we report modelled rephasing, non-rephasing and total correlation function spectra at a population waiting time of 50\,fs, in which excitons on the polymer and highly delocalised polarons are coupled this way, resulting in rapid population of delocalised polarons over ultrafast timescales. This is manifested by off-diagonal signal at $\sim100$\,meV above the diagonal in the total correlation spectrum, as observed in the experiment. Although this 2D spectral lineshape requires further investigation, these preliminary calculations show that our measurements are consistent with the process of noise-induced resonant charge tunnelling between intra-chain excitons and delocalized CT states proposed by some of us~\cite{BittnerNatComms2014}. 

Photocurrent excitation spectroscopy is a sensitive and broadly useful tool to study the nature of photoexcitations in semiconductors. Moreover, from an application point of view, it is an essential technique to unravel the charge generation process in view of the optimization of solar cells based on novel materials. It is particularly useful for materials in which photocarriers are not produced directly by inter-band transitions as in bulk semiconductors, but in which precursors to photocarriers, such as excitons in molecular semiconductors, can be probed via their excitation spectral lineshape. In general, 2D electronic spectroscopies permit identification of homogeneous and inhomogeneous spectral lineshape contributions, and because they are nonlinear optical techniques, they can reveal microscopic interchromophore couplings through the presence of cross peaks (off-diagonal signal) in the 2D correlation spectrum~\cite{ChoChemRev2008}. The 2DPC spectra reported in Fig.~\ref{fgr:2DPC} allow one to not only to discern the photocurrent excitation spectrum as in linear photocurrent action measurements, but also to identify spectral correlations. To our knowledge, there are two published examples of this approach in the current literature -- Nardin~et~al. demonstrated its application towards photocurrent excitation measurements in GaAs quantum wells produced for 2D electron-gas studies\cite{NardinOE2013}, and Karki~et~al. explored carrier multiplication processes in PbS solar cells based on colloidal semiconductor quantum dots \cite{KarkiNatComms2014}. 
The remarkably good quality of the signal-to-noise ratio in our data constitutes a promising step towards the employment of this experimental technique for the investigation of the photocharge generation process in OPVs and related excitonic solar-cells and photodetectors.
\section{Conclusions}
In this work we presented 2D coherent photocurrent excitation spectra measured on polymer-based solar cell. Our data show how this novel technique can be applied in the investigation of the charge photogeneration process in such systems. The unique added value of this new experimental method stems from the possibility to directly probe the final observable of interest, i.e. photocurrent, for the understanding of the the mechanism of primary charge generation in organic solar cells in their operating regime.
\begin{acknowledgements}
We are indebted to Prof.\ Andy Marcus and Dr.\ Julia Widom for their assistance in the construction of the apparatus. CS acknowledges NSERC, FRQNT, the CRC in Organic Semiconductor Materials, and the UdeM Research Chair.
\end{acknowledgements}
\bibliography{pss}

\begin{thebibliography}{14}%
\makeatletter
\providecommand \@ifxundefined [1]{%
 \@ifx{#1\undefined}
}%
\providecommand \@ifnum [1]{%
 \ifnum #1\expandafter \@firstoftwo
 \else \expandafter \@secondoftwo
 \fi
}%
\providecommand \@ifx [1]{%
 \ifx #1\expandafter \@firstoftwo
 \else \expandafter \@secondoftwo
 \fi
}%
\providecommand \natexlab [1]{#1}%
\providecommand \enquote  [1]{``#1''}%
\providecommand \bibnamefont  [1]{#1}%
\providecommand \bibfnamefont [1]{#1}%
\providecommand \citenamefont [1]{#1}%
\providecommand \href@noop [0]{\@secondoftwo}%
\providecommand \href [0]{\begingroup \@sanitize@url \@href}%
\providecommand \@href[1]{\@@startlink{#1}\@@href}%
\providecommand \@@href[1]{\endgroup#1\@@endlink}%
\providecommand \@sanitize@url [0]{\catcode `\\12\catcode `\$12\catcode
  `\&12\catcode `\#12\catcode `\^12\catcode `\_12\catcode `\%12\relax}%
\providecommand \@@startlink[1]{}%
\providecommand \@@endlink[0]{}%
\providecommand \url  [0]{\begingroup\@sanitize@url \@url }%
\providecommand \@url [1]{\endgroup\@href {#1}{\urlprefix }}%
\providecommand \urlprefix  [0]{URL }%
\providecommand \Eprint [0]{\href }%
\providecommand \doibase [0]{http://dx.doi.org/}%
\providecommand \selectlanguage [0]{\@gobble}%
\providecommand \bibinfo  [0]{\@secondoftwo}%
\providecommand \bibfield  [0]{\@secondoftwo}%
\providecommand \translation [1]{[#1]}%
\providecommand \BibitemOpen [0]{}%
\providecommand \bibitemStop [0]{}%
\providecommand \bibitemNoStop [0]{.\EOS\space}%
\providecommand \EOS [0]{\spacefactor3000\relax}%
\providecommand \BibitemShut  [1]{\csname bibitem#1\endcsname}%
\let\auto@bib@innerbib\@empty
%</preamble>
\bibitem [{\citenamefont {Halls}\ \emph {et~al.}(1995)\citenamefont {Halls},
  \citenamefont {Walsh}, \citenamefont {Greenham}, \citenamefont {Marseglia},
  \citenamefont {Friend}, \citenamefont {Moratti},\ and\ \citenamefont
  {Holmes}}]{Halls:1995uq}%
  \BibitemOpen
  \bibfield  {author} {\bibinfo {author} {\bibfnamefont {J.~J.~M.}\
  \bibnamefont {Halls}}, \bibinfo {author} {\bibfnamefont {C.~A.}\ \bibnamefont
  {Walsh}}, \bibinfo {author} {\bibfnamefont {N.~C.}\ \bibnamefont {Greenham}},
  \bibinfo {author} {\bibfnamefont {E.~A.}\ \bibnamefont {Marseglia}}, \bibinfo
  {author} {\bibfnamefont {R.~H.}\ \bibnamefont {Friend}}, \bibinfo {author}
  {\bibfnamefont {S.~C.}\ \bibnamefont {Moratti}}, \ and\ \bibinfo {author}
  {\bibfnamefont {A.~B.}\ \bibnamefont {Holmes}},\ }\href
  {http://dx.doi.org/10.1038/376498a0} {\bibfield  {journal} {\bibinfo
  {journal} {Nature}\ }\textbf {\bibinfo {volume} {376}},\ \bibinfo {pages}
  {498} (\bibinfo {year} {1995})}\BibitemShut {NoStop}%
\bibitem [{\citenamefont {G\'elinas}\ \emph {et~al.}(2011)\citenamefont
  {G\'elinas}, \citenamefont {Par\'e-Labrosse}, \citenamefont {Brosseau},
  \citenamefont {Albert-Seifried}, \citenamefont {Mcneill}, \citenamefont
  {Kirov}, \citenamefont {Howard}, \citenamefont {Leonelli}, \citenamefont
  {Friend},\ and\ \citenamefont {Silva}}]{Gelinas:2011vn}%
  \BibitemOpen
  \bibfield  {author} {\bibinfo {author} {\bibfnamefont {S.}~\bibnamefont
  {G\'elinas}}, \bibinfo {author} {\bibfnamefont {O.}~\bibnamefont
  {Par\'e-Labrosse}}, \bibinfo {author} {\bibfnamefont {C.-N.}\ \bibnamefont
  {Brosseau}}, \bibinfo {author} {\bibfnamefont {S.}~\bibnamefont
  {Albert-Seifried}}, \bibinfo {author} {\bibfnamefont {C.~R.}\ \bibnamefont
  {Mcneill}}, \bibinfo {author} {\bibfnamefont {K.~R.}\ \bibnamefont {Kirov}},
  \bibinfo {author} {\bibfnamefont {I.~A.}\ \bibnamefont {Howard}}, \bibinfo
  {author} {\bibfnamefont {R.}~\bibnamefont {Leonelli}}, \bibinfo {author}
  {\bibfnamefont {R.~H.}\ \bibnamefont {Friend}}, \ and\ \bibinfo {author}
  {\bibfnamefont {C.}~\bibnamefont {Silva}},\ }\href {\doibase
  10.1021/jp200466y} {\bibfield  {journal} {\bibinfo  {journal} {Journal Of
  Physical Chemistry C}\ }\textbf {\bibinfo {volume} {115}},\ \bibinfo {pages}
  {7114} (\bibinfo {year} {2011})}\BibitemShut {NoStop}%
\bibitem [{\citenamefont {Park}\ \emph {et~al.}(2009)\citenamefont {Park},
  \citenamefont {Roy}, \citenamefont {Beaupre}, \citenamefont {Cho},
  \citenamefont {Coates}, \citenamefont {Moon}, \citenamefont {Moses},
  \citenamefont {Leclerc}, \citenamefont {Lee},\ and\ \citenamefont
  {Heeger}}]{Park:2009ys}%
  \BibitemOpen
  \bibfield  {author} {\bibinfo {author} {\bibfnamefont {S.~H.}\ \bibnamefont
  {Park}}, \bibinfo {author} {\bibfnamefont {A.}~\bibnamefont {Roy}}, \bibinfo
  {author} {\bibfnamefont {S.}~\bibnamefont {Beaupre}}, \bibinfo {author}
  {\bibfnamefont {S.}~\bibnamefont {Cho}}, \bibinfo {author} {\bibfnamefont
  {N.}~\bibnamefont {Coates}}, \bibinfo {author} {\bibfnamefont {J.~S.}\
  \bibnamefont {Moon}}, \bibinfo {author} {\bibfnamefont {D.}~\bibnamefont
  {Moses}}, \bibinfo {author} {\bibfnamefont {M.}~\bibnamefont {Leclerc}},
  \bibinfo {author} {\bibfnamefont {K.}~\bibnamefont {Lee}}, \ and\ \bibinfo
  {author} {\bibfnamefont {A.~J.}\ \bibnamefont {Heeger}},\ }\href {\doibase
  10.1038/NPHOTON.2009.69} {\bibfield  {journal} {\bibinfo  {journal} {Nature
  Photonics}\ }\textbf {\bibinfo {volume} {3}},\ \bibinfo {pages} {297}
  (\bibinfo {year} {2009})}\BibitemShut {NoStop}%
\bibitem [{\citenamefont {Falke}\ \emph {et~al.}(2014)\citenamefont {Falke},
  \citenamefont {Rozzi}, \citenamefont {Brida}, \citenamefont {Maiuri},
  \citenamefont {Amato}, \citenamefont {Sommer}, \citenamefont {De~Sio},
  \citenamefont {Rubio}, \citenamefont {Cerullo}, \citenamefont {Molinari},\
  and\ \citenamefont {Lienau}}]{FalkeScience2014}%
  \BibitemOpen
  \bibfield  {author} {\bibinfo {author} {\bibfnamefont {S.~M.}\ \bibnamefont
  {Falke}}, \bibinfo {author} {\bibfnamefont {C.~A.}\ \bibnamefont {Rozzi}},
  \bibinfo {author} {\bibfnamefont {D.}~\bibnamefont {Brida}}, \bibinfo
  {author} {\bibfnamefont {M.}~\bibnamefont {Maiuri}}, \bibinfo {author}
  {\bibfnamefont {M.}~\bibnamefont {Amato}}, \bibinfo {author} {\bibfnamefont
  {E.}~\bibnamefont {Sommer}}, \bibinfo {author} {\bibfnamefont
  {A.}~\bibnamefont {De~Sio}}, \bibinfo {author} {\bibfnamefont
  {A.}~\bibnamefont {Rubio}}, \bibinfo {author} {\bibfnamefont
  {G.}~\bibnamefont {Cerullo}}, \bibinfo {author} {\bibfnamefont
  {E.}~\bibnamefont {Molinari}}, \ and\ \bibinfo {author} {\bibfnamefont
  {C.}~\bibnamefont {Lienau}},\ }\href {\doibase 10.1126/science.1249771}
  {\bibfield  {journal} {\bibinfo  {journal} {Science}\ }\textbf {\bibinfo
  {volume} {344}},\ \bibinfo {pages} {1001} (\bibinfo {year} {2014})},\ \Eprint
  {http://arxiv.org/abs/http://www.sciencemag.org/content/344/6187/1001.full.pdf}
  {http://www.sciencemag.org/content/344/6187/1001.full.pdf} \BibitemShut
  {NoStop}%
\bibitem [{\citenamefont {Vandewal}\ \emph {et~al.}(2014)\citenamefont
  {Vandewal}, \citenamefont {Albrecht}, \citenamefont {Hoke}, \citenamefont
  {Graham}, \citenamefont {Widmer}, \citenamefont {Douglas}, \citenamefont
  {Schubert}, \citenamefont {Mateker}, \citenamefont {Bloking}, \citenamefont
  {Burkhard}, \citenamefont {Sellinger}, \citenamefont {Fr{\'e}chet},
  \citenamefont {Amassian}, \citenamefont {Riede}, \citenamefont {McGehee},
  \citenamefont {Neher},\ and\ \citenamefont {Salleo}}]{VandewalNatMat2014}%
  \BibitemOpen
  \bibfield  {author} {\bibinfo {author} {\bibfnamefont {K.}~\bibnamefont
  {Vandewal}}, \bibinfo {author} {\bibfnamefont {S.}~\bibnamefont {Albrecht}},
  \bibinfo {author} {\bibfnamefont {E.~T.}\ \bibnamefont {Hoke}}, \bibinfo
  {author} {\bibfnamefont {K.~R.}\ \bibnamefont {Graham}}, \bibinfo {author}
  {\bibfnamefont {J.}~\bibnamefont {Widmer}}, \bibinfo {author} {\bibfnamefont
  {J.~D.}\ \bibnamefont {Douglas}}, \bibinfo {author} {\bibfnamefont
  {M.}~\bibnamefont {Schubert}}, \bibinfo {author} {\bibfnamefont {W.~R.}\
  \bibnamefont {Mateker}}, \bibinfo {author} {\bibfnamefont {J.~T.}\
  \bibnamefont {Bloking}}, \bibinfo {author} {\bibfnamefont {G.~F.}\
  \bibnamefont {Burkhard}}, \bibinfo {author} {\bibfnamefont {A.}~\bibnamefont
  {Sellinger}}, \bibinfo {author} {\bibfnamefont {J.~M.~J.}\ \bibnamefont
  {Fr{\'e}chet}}, \bibinfo {author} {\bibfnamefont {A.}~\bibnamefont
  {Amassian}}, \bibinfo {author} {\bibfnamefont {M.~K.}\ \bibnamefont {Riede}},
  \bibinfo {author} {\bibfnamefont {M.~D.}\ \bibnamefont {McGehee}}, \bibinfo
  {author} {\bibfnamefont {D.}~\bibnamefont {Neher}}, \ and\ \bibinfo {author}
  {\bibfnamefont {A.}~\bibnamefont {Salleo}},\ }\href
  {http://dx.doi.org/10.1038/nmat3807} {\bibfield  {journal} {\bibinfo
  {journal} {Nat Mater}\ }\textbf {\bibinfo {volume} {13}},\ \bibinfo {pages}
  {63} (\bibinfo {year} {2014})}\BibitemShut {NoStop}%
\bibitem [{\citenamefont {G{\'e}linas}\ \emph {et~al.}(2014)\citenamefont
  {G{\'e}linas}, \citenamefont {Rao}, \citenamefont {Kumar}, \citenamefont
  {Smith}, \citenamefont {Chin}, \citenamefont {Clark}, \citenamefont {van~der
  Poll}, \citenamefont {Bazan},\ and\ \citenamefont
  {Friend}}]{GelinasScience2014}%
  \BibitemOpen
  \bibfield  {author} {\bibinfo {author} {\bibfnamefont {S.}~\bibnamefont
  {G{\'e}linas}}, \bibinfo {author} {\bibfnamefont {A.}~\bibnamefont {Rao}},
  \bibinfo {author} {\bibfnamefont {A.}~\bibnamefont {Kumar}}, \bibinfo
  {author} {\bibfnamefont {S.~L.}\ \bibnamefont {Smith}}, \bibinfo {author}
  {\bibfnamefont {A.~W.}\ \bibnamefont {Chin}}, \bibinfo {author}
  {\bibfnamefont {J.}~\bibnamefont {Clark}}, \bibinfo {author} {\bibfnamefont
  {T.~S.}\ \bibnamefont {van~der Poll}}, \bibinfo {author} {\bibfnamefont
  {G.~C.}\ \bibnamefont {Bazan}}, \ and\ \bibinfo {author} {\bibfnamefont
  {R.~H.}\ \bibnamefont {Friend}},\ }\href {\doibase 10.1126/science.1246249}
  {\bibfield  {journal} {\bibinfo  {journal} {Science}\ }\textbf {\bibinfo
  {volume} {343}},\ \bibinfo {pages} {512} (\bibinfo {year}
  {2014})}\BibitemShut {NoStop}%
\bibitem [{\citenamefont {Provencher}\ \emph {et~al.}(2014)\citenamefont
  {Provencher}, \citenamefont {Berub{\'e}}, \citenamefont {Parker},
  \citenamefont {Greetham}, \citenamefont {Hellmann}, \citenamefont
  {C{\^o}t{\'e}}, \citenamefont {Stingelin}, \citenamefont {Silva},\ and\
  \citenamefont {Hayes}}]{ProvencherNatComms2014}%
  \BibitemOpen
  \bibfield  {author} {\bibinfo {author} {\bibfnamefont {F.}~\bibnamefont
  {Provencher}}, \bibinfo {author} {\bibfnamefont {N.}~\bibnamefont
  {Berub{\'e}}}, \bibinfo {author} {\bibfnamefont {A.~W.}\ \bibnamefont
  {Parker}}, \bibinfo {author} {\bibfnamefont {G.~M.}\ \bibnamefont
  {Greetham}}, \bibinfo {author} {\bibfnamefont {C.}~\bibnamefont {Hellmann}},
  \bibinfo {author} {\bibfnamefont {M.}~\bibnamefont {C{\^o}t{\'e}}}, \bibinfo
  {author} {\bibfnamefont {N.}~\bibnamefont {Stingelin}}, \bibinfo {author}
  {\bibfnamefont {C.}~\bibnamefont {Silva}}, \ and\ \bibinfo {author}
  {\bibfnamefont {S.~C.}\ \bibnamefont {Hayes}},\ }\href {\doibase
  10.1038/ncomms5288} {\bibfield  {journal} {\bibinfo  {journal} {Nat.
  Commun.}\ }\textbf {\bibinfo {volume} {5}},\ \bibinfo {pages} {428801}
  (\bibinfo {year} {2014})}\BibitemShut {NoStop}%
\bibitem [{\citenamefont {Clarke}\ and\ \citenamefont
  {Durrant}(2010)}]{Clarke:2010fk}%
  \BibitemOpen
  \bibfield  {author} {\bibinfo {author} {\bibfnamefont {T.~M.}\ \bibnamefont
  {Clarke}}\ and\ \bibinfo {author} {\bibfnamefont {J.~R.}\ \bibnamefont
  {Durrant}},\ }\href {\doibase 10.1021/cr900271s} {\bibfield  {journal}
  {\bibinfo  {journal} {Chem. Rev.}\ }\textbf {\bibinfo {volume} {110}},\
  \bibinfo {pages} {6736} (\bibinfo {year} {2010})}\BibitemShut {NoStop}%
\bibitem [{\citenamefont {Bittner}\ and\ \citenamefont
  {Silva}(2014)}]{BittnerNatComms2014}%
  \BibitemOpen
  \bibfield  {author} {\bibinfo {author} {\bibfnamefont {E.~R.}\ \bibnamefont
  {Bittner}}\ and\ \bibinfo {author} {\bibfnamefont {C.}~\bibnamefont
  {Silva}},\ }\href {\doibase 10.1038/ncomms4119} {\bibfield  {journal}
  {\bibinfo  {journal} {Nat. Commun.}\ }\textbf {\bibinfo {volume} {5}},\
  \bibinfo {pages} {311901} (\bibinfo {year} {2014})}\BibitemShut {NoStop}%
\bibitem [{\citenamefont {Tamura}\ and\ \citenamefont
  {Burghardt}(2014)}]{TamuraJACS2014}%
  \BibitemOpen
  \bibfield  {author} {\bibinfo {author} {\bibfnamefont {H.}~\bibnamefont
  {Tamura}}\ and\ \bibinfo {author} {\bibfnamefont {I.}~\bibnamefont
  {Burghardt}},\ }\href {\doibase 10.1021/ja4093874} {\bibfield  {journal}
  {\bibinfo  {journal} {J.~Am. Chem. Soc.}\ }\textbf {\bibinfo {volume}
  {135}},\ \bibinfo {pages} {16364} (\bibinfo {year} {2014})}\BibitemShut
  {NoStop}%
\bibitem [{\citenamefont {Tekavec}\ \emph {et~al.}(2007)\citenamefont
  {Tekavec}, \citenamefont {Lott},\ and\ \citenamefont
  {Marcus}}]{TekavecJCP2007}%
  \BibitemOpen
  \bibfield  {author} {\bibinfo {author} {\bibfnamefont {P.~F.}\ \bibnamefont
  {Tekavec}}, \bibinfo {author} {\bibfnamefont {G.~A.}\ \bibnamefont {Lott}}, \
  and\ \bibinfo {author} {\bibfnamefont {A.~H.}\ \bibnamefont {Marcus}},\
  }\href {\doibase 10.1063/1.280056} {\bibfield  {journal} {\bibinfo  {journal}
  {J. Chem Phys.}\ }\textbf {\bibinfo {volume} {127}},\ \bibinfo {pages}
  {214307} (\bibinfo {year} {2007})}\BibitemShut {NoStop}%
\bibitem [{\citenamefont {Nardin}\ \emph {et~al.}(2013)\citenamefont {Nardin},
  \citenamefont {Autry}, \citenamefont {Silverman},\ and\ \citenamefont
  {Cundiff}}]{NardinOE2013}%
  \BibitemOpen
  \bibfield  {author} {\bibinfo {author} {\bibfnamefont {G.}~\bibnamefont
  {Nardin}}, \bibinfo {author} {\bibfnamefont {T.~M.}\ \bibnamefont {Autry}},
  \bibinfo {author} {\bibfnamefont {K.~L.}\ \bibnamefont {Silverman}}, \ and\
  \bibinfo {author} {\bibfnamefont {S.~T.}\ \bibnamefont {Cundiff}},\ }\href
  {\doibase 10.1364/OE.21.028617} {\bibfield  {journal} {\bibinfo  {journal}
  {Opt. Express}\ }\textbf {\bibinfo {volume} {21}},\ \bibinfo {pages} {28617}
  (\bibinfo {year} {2013})}\BibitemShut {NoStop}%
\bibitem [{\citenamefont {Cho}(2008)}]{ChoChemRev2008}%
  \BibitemOpen
  \bibfield  {author} {\bibinfo {author} {\bibfnamefont {M.}~\bibnamefont
  {Cho}},\ }\href@noop {} {\bibfield  {journal} {\bibinfo  {journal} {Chemical
  Reviews}\ }\textbf {\bibinfo {volume} {108}},\ \bibinfo {pages} {1331}
  (\bibinfo {year} {2008})}\BibitemShut {NoStop}%
\bibitem [{\citenamefont {Karki}\ \emph {et~al.}(2014)\citenamefont {Karki},
  \citenamefont {Widom}, \citenamefont {Seibt}, \citenamefont {Moody},
  \citenamefont {Lonergan}, \citenamefont {Pullerits},\ and\ \citenamefont
  {Marcus}}]{KarkiNatComms2014}%
  \BibitemOpen
  \bibfield  {author} {\bibinfo {author} {\bibfnamefont {K.~J.}\ \bibnamefont
  {Karki}}, \bibinfo {author} {\bibfnamefont {J.~R.}\ \bibnamefont {Widom}},
  \bibinfo {author} {\bibfnamefont {J.}~\bibnamefont {Seibt}}, \bibinfo
  {author} {\bibfnamefont {I.}~\bibnamefont {Moody}}, \bibinfo {author}
  {\bibfnamefont {M.~C.}\ \bibnamefont {Lonergan}}, \bibinfo {author}
  {\bibfnamefont {T.}~\bibnamefont {Pullerits}}, \ and\ \bibinfo {author}
  {\bibfnamefont {A.~H.}\ \bibnamefont {Marcus}},\ }\href {\doibase
  10.1038/ncomms6869} {\bibfield  {journal} {\bibinfo  {journal} {Nat.
  Commun.}\ }\textbf {\bibinfo {volume} {5}},\ \bibinfo {pages} {5869}
  (\bibinfo {year} {2014})}\BibitemShut {NoStop}%
\end{thebibliography}

\end{document}